\documentclass[10pt,twocolumn,letterpaper]{article}

\usepackage{cvpr}              %
\usepackage{multirow}
\usepackage{colortbl}
\usepackage{graphicx}
\definecolor{almond}{rgb}{0.94, 0.87, 0.8}

\usepackage{lipsum}

\usepackage{stfloats}

\definecolor{cvprblue}{rgb}{0.21,0.49,0.74}
\usepackage[pagebackref,breaklinks,colorlinks,allcolors=cvprblue]{hyperref}

\def\paperID{13457} %
\def\confName{CVPR}
\def\confYear{2025}

\title{CAV-MAE Sync: Improving Contrastive Audio-Visual Mask Autoencoders via Fine-Grained Alignment}

\author{Edson Araujo$^1$\thanks{\small{\texttt{araujo@em.uni-frankfurt.de}}}\quad Andrew Rouditchenko$^2$\quad Yuan Gong$^2$\quad Saurabhchand Bhati$^2$\\
Samuel Thomas$^3$\quad Brian Kingsbury$^3$\quad Leonid Karlinsky$^3$\quad Rogerio Feris$^{3,4}$
\\ James R. Glass$^2$\quad Hilde Kuehne$^{1,4,5}$ \\
\small{$^1$Goethe University of Frankfurt, $^2$MIT, $^3$IBM Research, $^4$MIT-IBM Watson AI Lab, $^5$Tuebingen AI Center/University of Tuebingen}
}

\begin{document}
\maketitle

\begin{abstract}

Recent advances in audio-visual learning have shown promising results in learning representations across modalities. However,  most approaches rely on global audio representations that fail to capture fine-grained temporal correspondences with visual frames.
Additionally, existing methods often struggle with 
conflicting optimization objectives when trying to jointly learn reconstruction and cross-modal alignment. 
In this work, we propose CAV-MAE Sync as a simple yet effective extension of the original CAV-MAE~\cite{gong2023cavmae}  framework for self-supervised audio-visual learning. 
We address three key challenges: First, we tackle the granularity mismatch between modalities by treating audio as a temporal sequence aligned with video frames, rather than using global representations. Second, we resolve conflicting optimization goals by separating contrastive and reconstruction objectives through dedicated global tokens. Third, we improve spatial localization by introducing learnable register tokens that reduce the semantic load on patch tokens. We evaluate the proposed approach on AudioSet, VGG Sound, and the ADE20K Sound dataset on zero-shot retrieval, classification, and localization tasks demonstrating state-of-the-art performance and outperforming more complex architectures. Code is available at \href{https://github.com/edsonroteia/cav-mae-sync}{https://github.com/edsonroteia/cav-mae-sync}.
\end{abstract}

\section{Introduction}
\label{sec:intro}
\begin{figure}[t!]
    \centering
    \includegraphics[width=\linewidth]{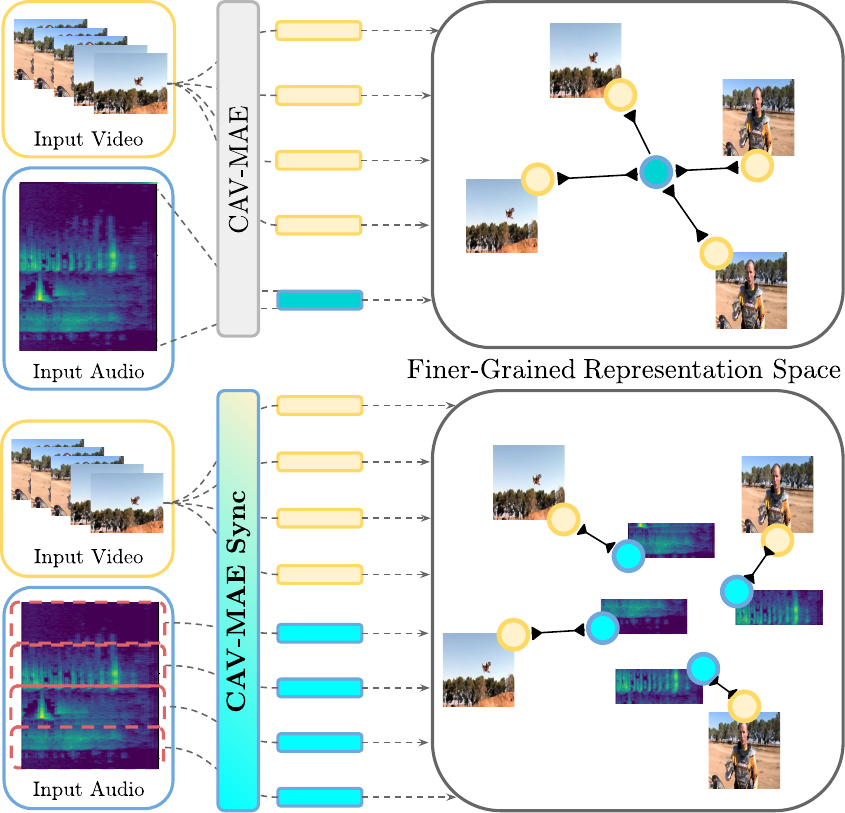}
    \caption{By representing audio with multiple finer-grained representations aligned with individual video frames, CAV-MAE Sync improves the precision of audio-visual alignment, in contrast to the original CAV-MAE, which uses a global audio representation that struggles with fine-grained temporal correspondence.}
    \label{fig:teaser}
\end{figure}

Humans perceive the world in a multimodal way where especially auditory and visual perception are very closely connected. %
As a result, jointly learning the representations of both modalities has been a longstanding active research topic in multimodal learning ~\cite{aytar2016soundnet,arandjelovic2017look, zhao2018sound,afouras2020self,cheng2020look,rouditchenko2021avlnet, lin2023lavish}. Specifically audio-visual alignment has been tackled from multiple perspectives, with major works focusing on contrastive learning \cite{ma2021active, chen2021distilling,sun2023learning}, but also exploring fusion-based methods \cite{hori2018multimodal, lee2020cross, ye2021temporal, senocak2023event}. Recently, multitask formulations combine multiple learning objectives and have emerged as a promising direction for audio-visual representation learning. In particular, CAV-MAE ~\cite{gong2023cavmae} introduced a framework that jointly optimizes contrastive alignment between modalities and masked reconstruction within each modality. By leveraging both cross-modal and intra-modal learning signals, this approach has emerged as a foundational architecture that has inspired several follow-up works \cite{huang2024mavil,lin2024avsiam,guo2024crossmae}.

We argue that while these methods have achieved impressive results, they show two significant limitations. First, most of them align the video and audio information based on a global audio representation, thus matching, for example, 10 seconds of audio to a single video frame.
Second, while the joint learning of cross-modal similarities together with in-modality reconstruction has proven to be a successful strategy, the vanilla implementation of this idea suffers from the fact that both objectives need to be achieved by a single representation, learned in a joint layer. This can be problematic as having a too similar representation for audio and vision can inhibit their reconstruction and vice versa.

To address this, we propose CAV-MAE Sync, a simple yet effective extension of the original CAV-MAE framework for audio-visual learning that takes advantage of the natural temporal alignment between modalities while at the same time relaxing the constraint of a joint representation. 
Namely, our work addresses three key challenges of this architecture as follows: 
\textit{i}) First, we tackle the granularity mismatch between visual and audio modalities. While existing methods typically operate on global audio-visual alignment objectives, we propose treating audio as a temporal sequence of instances, better matching the inherent structure of both modalities. This process is illustrated in Figure \ref{fig:teaser}.
\textit{ii}) Second, we resolve the tension between contrastive and reconstruction objectives. Current approaches like CAV-MAE simultaneously enforce representation similarity through contrastive loss while applying reconstruction loss on the same embedding layer. We argue this creates conflicting optimization goals. Our solution introduces separate global tokens, allowing each objective to operate in its optimal space.
\textit{iii}) Third, we introduce learnable register tokens into the pipeline. Similar to the results seen in \cite{darcet2024registers}, these registers further alleviate the semantic load of the patch tokens, while also allowing for finer-grained audio-visual alignment and thus better localization.

To evaluate our approach, we conduct experiments on zero-shot audio-visual retrieval, linear probing for audio-visual classification, and localization on the well-known VGGSound~\cite{chen2020vggsound}, AudioSet~\cite{gemmeke2017audio}, and ADE20K~\cite{zhou2017scene} datasets. Our results show that the proposed CAV-MAE Sync is not only superior to the original CAV-MAE architecture but also competes with significantly more complex architectures and achieves state-of-the-art performance on all tasks.

\noindent Our contributions can be summarized as follows:
(1) We propose CAV-MAE Sync, an extension of the CAV-MAE architecture that allows for a fine-grained temporal resolution on the audio side to support direct vision-audio alignment.
(2) We introduce global tokens to disentangle the inhibiting contrastive and reconstruction objectives and add registers to the pipeline to further de-noise the ViT signal.
(3) We evaluate the proposed setup on various downstream tasks and show a superior performance, even compared to significantly more complex architectures.

\section{Related Work}
\label{sec:rw}

\noindent\textbf{Contrastive Audio-Visual Masked Autoencoder Models. }
CAV-MAE \cite{gong2023cavmae} presented the first audio-visual model that leverages both contrastive learning and masked autoencoder objectives, and together with AVMAE \cite{georgescu2023audiovisual}, pioneered the self-supervised objective of masked autoencoding in the audio-visual domain. By combining the two learning tasks, CAV-MAE demonstrated strong performance across audio-visual tasks, effectively learning representations that capture both modality-specific features and cross-modal relationships.
Building on CAV-MAE's success, several works have proposed improvements to this contrastive audio-visual masked autoencoder framework. CrossMAE \cite{guo2024crossmae} introduced a region-aware approach using dual encoders and a fusion module to predict masked regions in both modalities, enabling fine-grained cross-modal understanding. MaViL \cite{huang2024mavil} built on top of the framework by performing video-level encoding and introducing a sophisticated masking strategy with both inter-modal contrasting between matched video-audio pairs and intra-modal contrasting between masked views of the same modality. AVSiam \cite{lin2024avsiam} proposed a parameter-efficient Siamese architecture that shares a single vision transformer backbone between modalities while maintaining the core CAV-MAE framework, reducing the model size while claiming to help bridge the modality gap between audio and visual representations. Unlike other CAV-MAE models, VAB \cite{su2024vab} focuses on masked audio token prediction in latent space using pre-trained tokenizers. While it can be fine-tuned with contrastive learning for retrieval tasks, its primary innovation is a visual-conditioned masked audio prediction that enables both representation learning and audio generation capabilities.
While these methods have made significant advances, they mostly operate on global audio-visual alignment objectives, missing opportunities for finer temporal granularity. Our work addresses this limitation by treating audio as a temporal sequence of instances, introducing separate global tokens to help disentangle competing objectives, and enhancing spatial localization through register tokens, achieving state-of-the-art performance with a simpler architecture.

\noindent\textbf{General Contrastive Audio-Visual Models. }
Initial approaches to audio-visual learning leveraged knowledge distillation \cite{aytar2016soundnet,owens2016ambient}, where well-trained visual models guided the optimization of audio networks. This was followed by the emergence of paired sample discrimination methods \cite{arandjelovic2017look,korbar2018cooperative,owens2018audio,arandjelovic2018objects}, which learned representations by distinguishing between matching and mismatched audio-visual pairs. Building on this successful paradigm, recent contrastive learning approaches \cite{morgado2021audio,sun2023learning,rouditchenko2021avlnet} have formalized the learning objective by maximizing similarity between positive pairs while minimizing similarity between negative examples in the embedding space. 
Several works have focused on improving contrastive learning through better sampling and data augmentation strategies. This includes active learning approaches for mining hard negatives \cite{ma2020active}, robust sampling to handle temporal misalignment \cite{morgado2021robust}, and multi-view techniques \cite{patrick2021compositions,wang2021multimodal,zeng2021contrastive,recasens2021broaden} that leverage both global and local temporal context. Recent work has explored making representations more robust by relaxing temporal synchronicity constraints \cite{sarkar2023self} and introducing equivariance \cite{kim2024equiav}. In another direction, several works have explored localization capabilities emerging from contrastive learning. Audio-Visual Correspondence \cite{arandjelovic2018objects} demonstrated that localization naturally emerges from the audio-visual correspondence task without explicit supervision. Building on this, Audio-Visual Associative Localizations \cite{harwath2018jointly} developed methods to explicitly link spoken audio descriptions with specific image regions. Chen et al. \cite{chen2021localizing} improved localization through hard negative mining in the contrastive learning process. 
Another line of work has further explored approaches to learning unified embeddings across multiple modalities. 
ImageBind \cite{girdhar2023imagebind} uses images as a semantic anchor, leveraging strong semantic relationships from image-text models to align multiple modalities like audio, depth, and thermal data. This enables zero-shot capabilities across modalities without requiring explicit pairings between all of them. Following, LanguageBind \cite{zhu2024languagebind} argues for directly aligning modalities to language, utilizing a pre-trained language encoder as the semantic anchor point, demonstrating good performance on language-related tasks through this direct alignment strategy. 
More recently, DenseAV \cite{hamilton2024separating} achieved strong localization by aligning dense audio-visual features through a dual encoder architecture built on top of Dino~\cite{caron2021emerging} and HuBERT~\cite{hsu2021hubert} backbones, showing good results on semantic segmentation and retrieval tasks.
Similar to DenseAV, we pursue fine-grained modal alignment but with a simpler approach that extends CAV-MAE with dedicated components for contrastive and reconstruction tasks.

\section{CAV-MAE Sync} \label{sec:method}
\begin{figure*}[t]
    \centering
    \includegraphics[width=\textwidth]{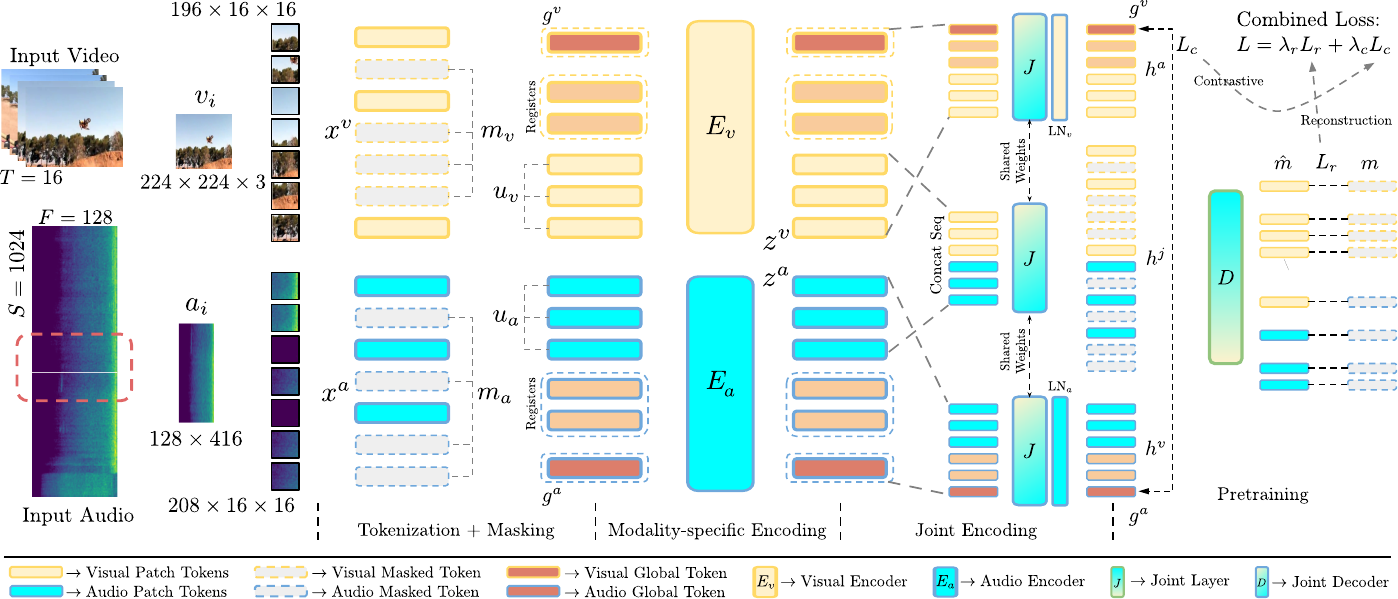}
    \caption{Overview of our approach. Our model processes video frames and audio segments in parallel through separate encoders $E_a$ and $E_v$, with the audio encoder $E_a$ operating on finer temporal granularity to better align with visual frames. Both modalities interact through the Joint Layer $L$ and the Joint Decoder $D$ The model is trained with both reconstruction and contrastive objectives.}
    \label{fig:model-overview}
\end{figure*}

\subsection{Overview}

Our method employs the contrastive masked autoencoder framework \cite{gong2023cavmae}, training the model to reconstruct both visual and audio signals while enhancing audio-visual alignment through a contrastive objective. Unlike traditional approaches that utilize a single audio representation, we implement a sequence of audio representations temporally aligned with visual frames. This strategy ensures more coherent temporal alignment between audio and visual modalities without complicating the model architecture. For downstream tasks, we leverage the finer-grained audio-visual correspondences learned during pretraining. Figure \ref{fig:model-overview} illustrates the data flow of our approach. In the following subsections, we first review the basics of CAV-MAE and then extend it in a second step toward the proposed CAV-MAE Sync framework.

\subsection{Background: CAV-MAE} \label{sec:background}

Given a video consisting of a set of frames and a respective audio signal, CAV-MAE uniformly samples frames from each video and selects for each training step one frame-audio pair consisting of a random frame $v_i$ and the visual representation of the full Mel spectrogram of the respective audio signal $a_i$. The respective two 2D inputs are then patchified. Then a portion of patches is randomly masked in each modality and a convolutional projection layer tokenizes the remaining frame and audio patches into sequences of visual and audio tokens respectively, also adding a modality type and a positional embedding.

The sequences of unmasked visual tokens $u_v$ and audio tokens $u_a$ are forwarded through separated encoders $E_v$ and $E_a$ to learn modality-specific representations $z^v$ and $z_a$. Note that, while both encoders share the same ViT architecture and are initialized from identical pretrained weights, they are trained independently without weight sharing.

After the modalities are encoded individually, their interactions are captured in a joint layer $J$, which is trained through three separate forward passes with shared transformer weights but individual layer normalizations, first one pass for the visual representations alone, second for the audio representations alone. The output patches of those two separate forward passes $h^a$ and $h^v$ are then averaged to form global representations of audio and visual modalities, $c_j^a$ and $c_i^v$, and used to compute the contrastive loss between the two modalities. The contrastive loss is then defined as:
\begin{equation}
L_c = -\frac{1}{N} \sum_{i=1}^{N} \log \left( \frac{\exp\left( s_{i,i} / \tau \right)}{\sum_{k\neq i} \exp\left( s_{i,k} / \tau \right) + \exp\left( s_{i,i} / \tau \right)} \right)
\end{equation}
with $ s_{i,j} = \|c_i^v\|^T \|c_j^a\| $ being the cosine similarity between normalized visual and audio representations, and $ \tau > 0 $ a temperature parameter of the similarity distribution.

Finally, for the third pass, the visual and audio tokens are concatenated into a single sequence. This joint representation is used for the masked autoencoding objective. The reconstructed patches are denoted as $\hat{m}^a_i$ for audio and $\hat{m}^v_i$ for video patches for each masked position $i$. The reconstruction terms $L^a_i$ and $L^v_i$ compute the mean squared error between predicted and original masked patches for audio and visual modalities, respectively:
\begin{align}
L^a_i = \frac{\sum_{i \in |m_a|} \left( \hat{m}_{a_i} - m_{a_i} \right)^2}{|m_a|} \\
L^v_i = \frac{\sum_{i \in |m_v|} \left( \hat{m}_{v_i} - m_{v_i} \right)^2}{|m_v|}
\end{align}
The reconstruction loss averages these terms over the batch:
\begin{equation}
L_r = \frac{1}{N} \sum_{i=1}^{N} \left( L^a_i + L^v_i \right)
\end{equation}
The final learning objective balances the contrastive and the reconstruction loss using weighting parameters $ \lambda_r $ and $ \lambda_c $ as $L = \lambda_c L_c + \lambda_r L_r$.
This weighted objective ensures that the model learns both modality-specific features through reconstruction and aligned cross-modal representations through contrastive learning.

\subsection{Improving Temporal Granularity} \label{sec:temporal_granularity}

We argue that the current contrastive matching of a full audio sequence to a single random frame is a rather loos contrastive objective, as 1) frames from different scenes will be mapped to the same audio as long as they come from the same video and 2) as longer audio usually also contain more than one audio class (e.g. in case of AudioSet) it not only maps several frames to the same audio but also to an audio encoding containing different classes. We therefore first aim to increase the temporal granularity to achieve a more precise audio-visual alignment. To this end, we extract audio segments corresponding to individual video frames. Unlike using a single audio spectrogram for the entire video, this method leverages the natural temporal alignment between audio and visual information and ensures that each audio segment is temporally aligned with its respective video frame, enhancing coherence between modalities.

\noindent\textbf{Temporal Alignment Process. }
Given a video with $T$ frames and its corresponding audio spectrogram of length $S$, we extract a fixed-length spectrogram segment of size $s_{\text{length}}$ for each frame $i$. Since video frames and audio spectrogram samples are extracted at different rates from the same time interval, we map each frame index to its corresponding position in the spectrogram using $s_{\text{center}_i} = \lfloor i \cdot S/T \rfloor$. We then extract a window centered at this position, adjusting the boundaries to handle edge cases. The segment extracted from the spectrogram is indexed from a starting position $s_{\text{start}_i}$ to an ending position $s_{\text{end}_i}$, where $s_{\text{start}_i} = s_{\text{center}_i} - \lfloor s_{\text{length}}/2 \rfloor$ and $s_{\text{end}_i} = s_{\text{start}_i} + s_{\text{length}}$.

\subsection{Disentangling Joint Modality Encoding} \label{sec:joint_modality_encoding}

In the original architecture \cite{gong2023cavmae}, patches are optimized for both contrastive and autoencoder objectives using a shallow joint layer, which can hinder the model's ability to learn distinct representations for each objective. To address this, we propose strategies to disentangle these objectives, enhancing the model's performance and representation quality.

\noindent\textbf{Global Token Integration. }
While traditional MAE approaches aggregate patch tokens to form global representations for downstream tasks \cite{gong2023cavmae, georgescu2023audiovisual}, we instead introduce dedicated global tokens for the contrastive objective. By separating the global representation pathway from the patch tokens, we reduce the information burden on patch tokens, which can now focus on reconstruction while the global tokens aggregate information during both single-modality and joint encoding stages.
We denote these global tokens as $ g^a $ and $ g^v $ for audio and visual modalities respectively. These tokens serve as global representations of their respective modalities and are used for the contrastive objective and downstream tasks. 
While these tokens are optimized primarily through the contrastive loss, the entire model's weights, including the encoders and joint layers, are still updated through backpropagation from both objectives.

\noindent\textbf{Register Tokens. }
Vision transformers often contain high-norm patch tokens that act as computation nodes rather than visual features in the self-supervised setting \cite{darcet2024registers}. We incorporate the idea of register tokens, which helps our method in two ways: It maintains the patch tokens dedicated to the reconstruction objective and allows the global tokens to focus on the contrastive objective. This disentanglement improves the model's ability to capture semantic information and perform localization. These register tokens are appended to $u_v$ and $u_a$ and are processed through the joint layer in the same manner as the global tokens.

\noindent\textbf{Adaptation of the Joint Layer. }
With the addition of global and register tokens, the joint layer is refined to handle modality-specific representations more effectively. The contrastive loss $ L_c $ now exclusively utilizes the global tokens ($ g^a $ and $ g^v $), computing similarity scores as $s_{i,j} = \|g_i^v\|^T \|g_j^a\|$.
This ensures that the contrastive objective operates on high-level modality representations, while the patch tokens remain to address the reconstruction objective.
By disentangling the contrastive and autoencoder objectives, the model can better optimize each task independently. The global tokens provide robust representations for cross-modal alignment, while the patch tokens focus on accurate reconstruction. This separation leads to improved performance in both representation learning and downstream tasks, as the model leverages specialized features for each objective without mutual interference.

\subsection{Downstream Tasks} \label{sec:downstream}

\subsubsection{Cross-Modal Retrieval}

For cross-modal retrieval, the goal is to retrieve relevant audio segments given a visual query and vice versa. Unlike approaches that use global tokens per modality, we leverage multiple temporal tokens to capture fine-grained relationships between audio and visual data.
For each video, we forward all frames and their corresponding audio segments through their respective encoders and joint layer. Using the definitions from Section \ref{sec:joint_modality_encoding}, we obtain the final global tokens $g^v$ and $g^a$ after passing through the joint layer $J$ with their corresponding layer normalizations $\text{LN}_v$ and $\text{LN}_a$.

\noindent\textbf{Similarity Calculation. }
For video-to-audio retrieval, consider a query video with a set of visual global tokens $V_q = \{g^v_1, ..., g^v_T\}$ and a target video with audio global tokens $A_t = \{g^a_1, ..., g^a_T\}$, with $T$ as number of temporal tokens per modality. We construct a similarity matrix $S = V_q A_t^\top$ between the two sets where each element $s_{i,j}$ represents the similarity between the $i$-th visual token of the query video and the $j$-th audio token of the target video.
We compute the final similarity score by averaging the diagonal of $S$, emphasizing the modality temporal alignment:
\vspace{-1mm}
\begin{equation}
\text{Similarity Score} = \frac{1}{T} \sum_{t=1}^{T} s_{t,t}
\vspace{-1mm}
\end{equation}
This diagonal-focused approach ensures that temporally corresponding token pairs contribute most strongly to the similarity score, promoting retrieval based on both semantic and temporal alignment. 
For a batch of videos, we compute the similarity scores between all query-target pairs to construct a ranking matrix $R$, where each element $R_{i,j}$ represents the similarity score between query video $i$ and target video $j$. The rankings are determined by sorting the similarity scores for each query, with higher scores indicating better matches.
\begin{figure}[t]
    \centering
    \includegraphics[width=\linewidth]{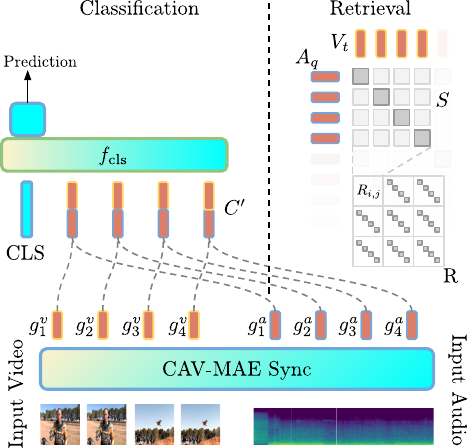}
    \caption{Illustration of our downstream tasks: (1) Classification: using CLS token with $f_{cls}$ projection for video-level prediction, and (2) Retrieval: computing similarity matrix R between audio query $A_q$ and video candidates $V_t$ for cross-modal matching.}
    \label{fig:downstream}
\end{figure}

Figure \ref{fig:downstream} illustrates both our classification and cross-modal retrieval tasks. 

\subsubsection{Classification}

For the classification task, we extend the sampling strategy to ensure comprehensive temporal coverage of each video. Instead of loading a single audio-visual segment per video instance, we sample all frames and their corresponding audio segments, effectively increasing the batch size to $B \cdot T$.

For classification, we first obtain the global tokens $g^v$ and $g^a$ from the visual and audio encoders for each temporal step $t$ in video $i$. These tokens are concatenated at each timestep to form a sequence $C_i$ of length $T$ containing the unified audio-visual representations.
We prepend a learnable CLS token to the sequence $C_i$ to aggregate temporal information, resulting in the final sequence $C'_i = \{\text{CLS}, C_{i,1}, \dots, C_{i,T}\}$ that is passed to the classification head.
Let $f_\text{cls}$ denote our classification head - a two-layer transformer followed by a linear projection. Given the extended sequence $C'_i$, the classification head produces predictions $\hat{y}_i = f_\text{cls}(C'_i)$. 
For multi-class tasks like AudioSet, we use binary cross-entropy loss. For single-class tasks like VGGSound, we use standard cross-entropy loss. In both cases, the model learns to aggregate temporal information over the concatenated audio-visual tokens.

\subsubsection{Sound-Prompted Semantic Segmentation}

For sound-prompted semantic segmentation, our goal is to make our model output a localization map in a frame given an audio query. 
We extract the global audio token $g^a$ and all visual tokens $h^v$ and compute the cosine similarity between each $h^v$ and $g^a$, forming the similarity matrix $L$ corresponding to the $14 \times 14$ patches grid of the frame. We then upscale this matrix to the original frame resolution of $224 \times 224$ pixels and use it as our predicted localization map.

\section{Evaluation} \label{sec:evaluation}
\begin{table*}[!ht]
\centering
\resizebox{\linewidth}{!}{
\begin{tabular}{@{}llcccccc cclccccccc@{}}
\toprule
\multirow{9}{*}{\rotatebox{90}{Visual $\rightarrow$ Audio}} & & \multicolumn{3}{c}{AudioSet Eval Subset} & \multicolumn{3}{c}{VGGSound Eval Subset} & \multirow{9}{*}{\rotatebox{90}{Audio $\rightarrow$ Visual}} & \multicolumn{3}{c}{AudioSet Eval Subset} & \multicolumn{3}{c}{VGGSound Eval Subset} \\ 
\cmidrule(lr){3-5} \cmidrule(lr){6-8} \cmidrule(lr){10-12} \cmidrule(lr){13-15}
 & \textbf{Baselines} &  R@1 & R@5 & R@10 & R@1 & R@5 & R@10 &  & R@1 & R@5 & R@10 & R@1 & R@5 & R@10 \\ 
\cmidrule(lr){2-8} \cmidrule(lr){10-15}
& \textcolor{lightgray}{VAB-Encodec \cite{su2024vab}} & \textcolor{lightgray}{$39.5$} & \textcolor{lightgray}{$65.4$} & \textcolor{lightgray}{$74.6$} & \textcolor{lightgray}{$33.5$} & \textcolor{lightgray}{$63.3$} & \textcolor{lightgray}{$74.3$} &  & \textcolor{lightgray}{$37.5$} & \textcolor{lightgray}{$64.0$} & \textcolor{lightgray}{$73.7$} & \textcolor{lightgray}{$34.9$} & \textcolor{lightgray}{$62.7$} & \textcolor{lightgray}{$73.1$} \\
\cmidrule(lr){2-8} \cmidrule(lr){10-15}
& CAV-MAE \cite{gong2023cavmae} & $16.1$ & $38.6$ & $49.3$ & $14.7$ & $35.3$ & $45.9$ &  & $9.5$ & $22.6$ & $32.4$ & $8.3$ & $23.8$ & $32.4$ \\
& CAV-MAE\textsuperscript{Scale+} \cite{gong2023cavmae} &  $18.8$ & $39.5$ & $50.1$ & $14.8$ & $34.2$ & $44.0$ &  &  $15.1$ & $34.0$ & $43.0$ & $12.8$ & $30.4$ & $40.3$ \\ 
& LanguageBind \cite{zhu2024languagebind} &  $6.4$ & $20.2$ & $28.3$ & $10.3$ & $30.1$ & $39.7$ &  &  $4.4$ & $15.0$ & $22.5$ & $6.5$ & $22.7$ & $33.5$ \\
& AVSiam \cite{lin2024avsiam} &  $19.7$ & $-$ & $-$ & $19.0$ & $-$ & $-$  &  &  $17.6$ & $-$ & $-$ & $20.4$ & $-$ & $-$ \\
& ImageBind \cite{girdhar2023imagebind} &  $22.1$ & $43.2$ & $52.6$ & $21.6$ & $43.4$ & $52.9$ &  &  $20.8$ & $42.6$ & $51.6$ & $20.7$ & $43.2$ & $53.4$ \\
\cmidrule(lr){2-8} \cmidrule(lr){10-15}
\rowcolor{almond}& \textbf{Ours} &  \textbf{35.2} & \textbf{58.3} & \textbf{67.6} & \textbf{27.9} & \textbf{51.7} & \textbf{61.8} &  & \textbf{27.9} & \textbf{52.4} & \textbf{62.2} & \textbf{23.2} & \textbf{46.2} & \textbf{58.1} \\ 
\bottomrule
\end{tabular}
}
\caption{Zero-shot retrieval results on AudioSet and VGGSound for Visual to Audio (V→A) and Audio to Visual (A→V) tasks. Our model achieves state-of-the-art zero-shot performance across all retrieval metrics (R@1, R@5, R@10) on both datasets, surpassing baselines like ImageBind and AVSiam. Fine-tuned VAB-Encodec scores are provided as an upper bound for comparison.}\label{tab:retrieval}
\end{table*}

\subsection{Datasets}

\noindent\textbf{AudioSet. } The full AudioSet-2M dataset contains over $2$ million $10$-second YouTube video clips annotated with $527$ audio event classes, which we use for pre-training. For downstream evaluation, we use AudioSet-20K \cite{gemmeke2017audio}, a balanced subset containing $20,000$ samples. For retrieval, we use the subsampled split provided by \cite{gong2023cavmae}.

\noindent\textbf{VGGSound. }The dataset \cite{chen2020vggsound} consists of $200,000$ $10$-second YouTube videos annotated with $309$ classes. Each video contains a visually evident sound source, verified through a pretrained vision classifier. This property makes VGGSound less noisy in terms of audio-visual correspondence.

\noindent\textbf{ADE20K\_Sound. } This dataset contains $106$ images from ADE20K \cite{zhou2017scene} paired with corresponding sound clips from VGGSound. The images and sounds span $20$ ADE20K classes, with each image containing objects that produce the paired sound (e.g., an image of a dog paired with a barking sound). The dataset was created by manually selecting ADE20K images and matching them with semantically relevant audio clips from VGGSound.

\subsection{Downstream Tasks}
To assess the capabilities of our proposed approach, we evaluate performance on three key downstream tasks. Further implementation details can be found in the supplement. 

\noindent\textbf{Zero-shot Audio-Visual Retrieval. } Given a query from one modality, the model must retrieve the corresponding sample from the other modality. We evaluate bidirectional retrieval (Visual→Audio and Audio→Visual) using Recall@k metrics ($k=\{1,5,10\}$) on AudioSet and VGGSound test sets. For fair comparison, we follow the evaluation protocol and subsampling from CAV-MAE \cite{gong2023cavmae}, using cosine similarity between embeddings to rank candidates.

\noindent\textbf{Sound-Prompted Image Segmentation. } Following \cite{hamilton2024separating}, we evaluate cross-modal localization using ADE20K\_Sound. Given an audio prompt, the model must segment corresponding regions in the paired image. Performance is measured via mean Average Precision (mAP) and mean Intersection over Union (mIoU) across $20$ classes.

\noindent\textbf{Classification. } We assess representation quality through linear probing on AudioSet and VGGSound classification tasks. Following standard practice, we freeze the pretrained encoder and train only a linear classifier, using mean Average Precision (mAP) for AudioSet's multi-label case and accuracy for VGGSound's single-label setting.

\subsection{Comparison to State-of-the-Art}

\noindent\textbf{Zero-shot Retrieval. }
We evaluate our model's retrieval performance in both directions - Visual to Audio (V→A) and Audio to Visual (A→V) - on AudioSet and VGGSound datasets, following the same subsampling strategy as \cite{gong2023cavmae}. 
For baselines, we compare against state-of-the-art audio-visual models including CAV-MAE \cite{gong2023cavmae}, ImageBind \cite{girdhar2023imagebind}, AVSiam \cite{lin2024avsiam}, and VAB \cite{su2024vab} as a reference for a model fine-tuned for retrieval serving as an upper bound. We report numbers from original papers and otherwise use officially released checkpoints where available.
The retrieval metrics are computed using cosine similarity between query and target embeddings as detailed in Section~\ref{sec:downstream}. As shown in Table~\ref{tab:retrieval}, our model achieves state-of-the-art performance in both directions, suggesting that our model learns a balanced joint embedding space. 
These results demonstrate that enforcing temporally consistent audio-visual correspondences during pretraining together with a disentangling of the contrastive MAE objective enables better generalization to downstream retrieval tasks.

\noindent\textbf{Classification. }
\begin{table}[t]
\footnotesize
\centering
\resizebox{\linewidth}{!}{
\begin{tabular}{@{}lcccc@{}}
\toprule
   \textbf{Baselines}                    & {Pretrain Dataset} & AS20K$\uparrow$ & VGGSound$\uparrow$  \\ \midrule
\textcolor{lightgray}{VAB-Encodec \cite{su2024vab}}           & \textcolor{lightgray}{AS-2M + VGGS}      & \textcolor{lightgray}{33.3}                 & \textcolor{lightgray}{57.6}               \\
\midrule
CAV-MAE \cite{gong2023cavmae}                                 & AS-2M      & $27.3$                  & -            \\
CAV-MAE\textsuperscript{Scale+} \cite{gong2023cavmae}           & AS-2M      & $28.5$               & $47.7$            \\
CAV-MAE\textsuperscript{Scale++} \cite{gong2023cavmae}          & AS-2M      & $29.2$               & $51.1$              \\
CAV-MAE\textsuperscript{Scale+++} \cite{gong2023cavmae}        & AS-2M      & $25.3$               & $51.6$               \\
MaViL \cite{huang2024mavil}                                  & AS-2M      & $30$                 & -               \\
\midrule
\rowcolor{almond} Ours                   & AS-2M      & \textbf{30.5}      & \textbf{52.7}              \\
\bottomrule
\end{tabular}
}
\vspace{-2mm}
\caption{Comparing audio-visual classification performance using linear probing. Numbers reported for AS20K are calculated using mAP (mean Average Precision) and VGGSound with accuracy.}
\label{tab:linear-probing}
\vspace{-5mm}
\end{table} 

While our model achieves strong retrieval performance through finer-grained temporal alignment, we also evaluate its representation learning capabilities through linear probing. Prior work has observed trade-offs between retrieval and representation learning performance, where optimizing for one task often comes at the expense of the other. 
Our goal is to maintain strong performance across all tasks through a unified architecture rather than specializing in a single objective.
Table \ref{tab:linear-probing} compares linear probing performance on AudioSet-20K and VGGSound classification. Our model achieves $30.5$ mAP on AudioSet and $52.7\%$ accuracy on VGGSound, outperforming CAV-MAE variants and MaViL when using only AudioSet-2M pretraining. 

\noindent\textbf{Sound-Prompted Image Segmentation. }
Following \cite{hamilton2024separating}, we finally assess our model's ability to localize sound sources in images using the ADE20K\_Sound dataset. This dataset pairs $106$ images from ADE20K with corresponding sound clips from VGGSound, spanning $20$ ADE20K classes. The task requires the model to segment regions in an image corresponding to a given sound prompt, effectively testing cross-modal capabilities. Performance is measured using mean Average Precision (mAP) and mean Intersection over Union (mIoU), averaged across the $20$ ADE20K classes. As shown in Table \ref{tab:localization}, our model achieves $22.7$ mIoU on sound-prompted segmentation, performing on par with previous self-supervised approaches like CAVMAE and ImageBind. Note that while the current best model, DenseAV, achieves higher performance ($24.2$ with DinoV2+LoRA), we only directly compare to approaches with identical or similar backbones and architecture. 
\begin{table}[t] 
    \centering
    \begin{tabular}{lcc}
        \toprule
        \textbf{Baselines} & \textbf{mAP} $\uparrow$ & \textbf{mIoU} $\uparrow$ \\
        \midrule
        DAVENet \cite{harwath2018jointly} & $16.8$ & $17.0$ \\
        CAVMAE \cite{gong2023cavmae} & $26.0^\dagger$ / $21.2$ & $20.5^\dagger$ / $20.9$ \\
        ImageBind \cite{girdhar2023imagebind} & $18.3$ & $19.1$ \\        
        \midrule
        \rowcolor{almond}Ours & \textbf{22.6} & \textbf{22.7} \\
        \bottomrule
    \end{tabular}
    \vspace{-2mm}
    \caption{Sound-prompted semantic segmentation: Comparison of sound localization methods on ADE20K Sound dataset \cite{hamilton2024separating}. $^\dagger$Original reported by \cite{hamilton2024separating} / our reproduction.}
    \label{tab:localization}
\end{table}

\subsection{Ablation Studies}
We conduct a series of ablation studies to evaluate the impact of different components on our model's performance. %
Tables~\ref{tab:retrieval_cavmae}--\ref{tab:ablation:masking} summarize the results of these experiments.

\begin{table}[t] \label{mainretrievaltable}
    \centering
    \begin{minipage}[c]{1\linewidth}
    \resizebox{\linewidth}{!}{
    \begin{tabular}{@{}lccccccc@{}}
    \toprule
    Visual $\rightarrow$ Audio & \multicolumn{3}{c}{AudioSet Eval Subset} & \multicolumn{3}{c}{VGGSound Eval Subset} \\ 
    \midrule
                                    & R@1           & R@5           & R@10          & R@1           & R@5           & R@10          \\ 
    \midrule
    CAV-MAE\textsuperscript{Scale+}     & $15.7$	& $35.2$	& $45.3$          & $11.1$          & $26.5$          & $35.0$          \\ 
    \multicolumn{7}{l}{\scriptsize\textit{$\hookrightarrow$ Increase batch size 128 $\rightarrow$ 256}} \\
    CAV-MAE\textsuperscript{Scale++}                  & $19.7$          & $40.2$          & $50.7$          & $14.8$          & $31.6$          & $41.0$          \\
    \multicolumn{7}{l}{\scriptsize\textit{$\hookrightarrow$ Pretrain and retrieve with 16 temporal tokens using diagonal similarity (Sec. \ref{sec:downstream})}} \\
    CAV-MAE\textsuperscript{Scale++}*     & $23.9$          & $46.8$          & $58.0$          & $16.1$          & $36.2$          & $46.1$          \\
    \multicolumn{7}{l}{\scriptsize\textit{$\hookrightarrow$ Increase batch size 256 $\rightarrow$ 512 and $\lambda_c$=0.1}} \\
    CAV-MAE\textsuperscript{Scale+++} ($\lambda_c$=0.1) & $30.1$          & $54.9$          & $64.5$          & $21.0$          & $44.0$          & $56.8$          \\
    \midrule
    \rowcolor{almond}\textbf{Ours} & \textbf{35.2} & \textbf{58.3} & \textbf{67.6} & \textbf{27.9} & \textbf{51.7} & \textbf{61.8}\\
    \bottomrule
    \end{tabular}
    }
    \end{minipage}\hfill
    \vspace{-2mm}
    \caption{Retrieval performance comparison showing the progression from CAV-MAE to our approach. The results demonstrate that simply using temporal tokens with diagonal similarity yields weaker performance than ours, highlighting the importance of our global and register tokens combined with fine-grained pretraining.}\label{tab:retrieval_cavmae}
    \vspace{-3mm}
\end{table}

\noindent\textbf{Establishing a Strong CAV-MAE Baseline. }
Table \ref{tab:retrieval_cavmae} shows our systematic optimization of CAV-MAE as a baseline. Starting with CAV-MAE\textsuperscript{Scale+}, increasing batch size to $256$ yields CAV-MAE\textsuperscript{Scale++} with improved AudioSet retrieval (R@1 from $18.8$ to $19.7$). Using $16$ temporal tokens during both pretraining and retrieval, along with diagonal sequence similarity for retrieval (CAV-MAE\textsuperscript{Scale++}*), further boosts performance to $23.9$ R@1. Finally, larger batches ($512$) and stronger contrastive weight ($\lambda_c=0.1$) in CAV-MAE\textsuperscript{Scale+++} achieves $30.1$ R@1. For fair comparison, our model uses the same hyperparameters with $4$s audio segments, reaching $35.2$ R@1 on AudioSet and $27.9$ R@1 on VGGSound with the same parameter count and a fraction of the token count.

\begin{figure}[ht]
    \footnotesize
    \includegraphics[width=\linewidth]{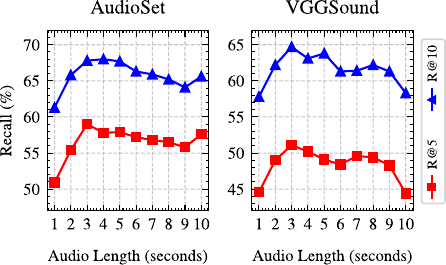}
    \vspace{-5mm}
    \caption{Impact of audio segment length on model performance. Experiments show $3$-$4$ second segments achieve optimal results while reducing computational costs compared to standard 10-second segments.} \label{tab:ablation:audio}
    \vspace{-3mm}
\end{figure}

\begin{table}[ht]
    \footnotesize
    
    \resizebox{\linewidth}{!}{
    \begin{tabular}{@{}lcccccccc@{}}
    \toprule
    \multicolumn{2}{c}{\# Frames} & \multicolumn{3}{c}{AudioSet Eval Subset} & \multicolumn{3}{c}{VGGSound Eval Subset} \\ 
    \cmidrule(lr){1-2} \cmidrule(lr){3-5} \cmidrule(lr){6-8}
    Train & Eval & R@1 & R@5 & R@10 & R@1 & R@5 & R@10 \\
    \midrule
    $10$ & $10$ & $31.2$ & $55.3$ & $65.6$ & $25.6$ & $49.6$ & $59.3$ \\
    $10$ & $16$ & $34$ & $57.3$ & $66.8$ & $27.0$ & $50.7$ & $60.6$ \\
    $16$ & $10$ & $32.3$ & $55.0$ & $65.3$ & $26.3$ & $48.7$ & $59.6$ \\
    \rowcolor{almond}$16$ & $16$ & \textbf{35.2} & \textbf{58.3} & \textbf{67.6} & \textbf{27.9} & \textbf{51.7} & \textbf{61.8} \\
    \bottomrule
    \end{tabular}}
    \vspace{-2mm}
    \caption{Effect of frame sampling density on retrieval performance during pre-training and evaluation. Results demonstrate consistent improvements with denser temporal sampling across both stages.} \label{tab:ablation:frames}
    \vspace{-3mm}
\end{table}

\begin{table}[ht]
    \footnotesize
    
    \resizebox{\linewidth}{!}{
    \begin{tabular}{@{}ccccccccc@{}}
    \toprule
    \multirow{2}{*}{\# Regs.} & \multicolumn{3}{c}{AudioSet Eval Subset} & \multicolumn{3}{c}{VGGSound Eval Subset} & \multicolumn{1}{c}{Localization} & \multicolumn{1}{c}{Classification}\\ 
    \cmidrule(lr){2-4} \cmidrule(lr){5-7} \cmidrule(lr){8-8} \cmidrule(lr){9-9}
    & R@1 & R@5 & R@10 & R@1 & R@5 & R@10 & mIoU & AS20k (mAP) \\
    \midrule
    $0$ & \textbf{35.5} & $59.0$ & $68.0$ & $27.9$ & $50.9$ & $62.4$ & $20.9$ & $27.1$\\    
    $4$ & $32.9$ & $58.1$ & $67.0$ & \textbf{28.2} & $51.3$ & $63.2$ & $21.7$ & $27.5$\\
    \rowcolor{almond}$8$ & $35.2$ & $58.3$ & $67.6$ & $27.9$ & $51.7$ & $61.8$  & $22.8$ & \textbf{30.8}\\
    $16$ & $34.9$ & \textbf{59.2} & \textbf{68.5} & $27.7$ & \textbf{52.2} & \textbf{64.0} & \textbf{22.9} & $26.8$\\
    \bottomrule
    \end{tabular}}
    \vspace{-2mm}
    \caption{Ablation studies on the number of registers. Results demonstrate consistent improvements with denser temporal sampling across both stages.} \label{tab:ablation:registers}
    \vspace{-3mm}
\end{table}

\begin{table}[ht]
    
    \resizebox{\linewidth}{!}{
    \begin{tabular}{@{}ccccccccc@{}}
    \toprule
    \multirow{2}{*}{Global} & \multicolumn{3}{c}{AudioSet Eval Subset} & \multicolumn{3}{c}{VGGSound Eval Subset} & \multicolumn{1}{c}{Localization} & \multicolumn{1}{c}{Classification}\\ 
    \cmidrule(lr){2-4} \cmidrule(lr){5-7} \cmidrule(lr){8-8} \cmidrule(lr){9-9}
    & R@1 & R@5 & R@10 & R@1 & R@5 & R@10 & mIoU & AS20k (mAP) \\
    \midrule
    w/o & $30.1$ & $55.3$ & $64.9$ & $21.6$ & $48.1$ & $60.7$ & $20.7$ & $26.6$ \\
     \rowcolor{almond}\textbf{w/} & \textbf{35.2} & \textbf{58.3} & \textbf{67.6} & \textbf{27.9} & \textbf{51.7} & \textbf{61.8} & \textbf{22.8} & \textbf{30.8}\\
    \bottomrule
    \end{tabular}}
    \vspace{-2mm}
    \caption{Effect of global token on model performance. The global token significantly improves cross-modal retrieval on both datasets while enhancing  localization and classification tasks, demonstrating its importance for capturing audio-visual relationships.} \label{tab:ablation:cls}
    \vspace{-3mm}
\end{table}

\begin{table}[ht]
    \footnotesize
    
    \resizebox{\linewidth}{!}{
    \begin{tabular}{@{}cccccccccc@{}}
    \toprule
    \multirow{2}{*}{Ratio} & \multicolumn{3}{c}{AudioSet Eval Subset} & \multicolumn{3}{c}{VGGSound Eval Subset} & \multicolumn{1}{c}{Localization} \\ 
    \cmidrule(lr){2-4} \cmidrule(lr){5-7} \cmidrule(lr){8-8}
    & R@1 & R@5 & R@10 & R@1 & R@5 & R@10 & mIoU \\
    \midrule
    0.6 & \textbf{39.1} & \textbf{63.3} & \textbf{71.7} & \textbf{28.6} & \textbf{55.5} & \textbf{68.7} & 19.1 \\
    \rowcolor{almond}0.75 & $35.2$ & $58.3$ & $67.6$ & $27.9$ & $51.7$ & $61.8$ & $22.8$ \\
    0.9 & $21.5$ & $42.3$ & $53.5$ & $16.7$ & $37.9$ & $50.1$ & $23.4$ \\
    \textit{multi \{0.6-0.9\}} & $27.3$ & $51.6$ & $62.1$ & $22.1$ & $45.7$ & $59.9$ & $21.2$ \\ 
    \bottomrule
    \end{tabular}}
    \vspace{-2mm}
    \caption{Effect of masking ratio on retrieval and localization performance. Trade-off analysis between different masking ratios shows $0.75$ provides optimal balance between tasks.} 
    \vspace{-3mm}
    \label{tab:ablation:masking}
\end{table}

\noindent\textbf{Audio Length Ablation. } To promote fine-grained temporal alignment between audio and visual modalities, we investigate the optimal temporal granularity for audio sampling, as videos typically contain multiple distinct audio events that should be precisely matched with their corresponding visual frames.  In Figure \ref{tab:ablation:audio}, our experiments show that $3$-$4$ second segments provide optimal performance for our architecture and pre-training approach, outperforming the standard 10-second segments in zero-shot retrieval tasks. Note that this shorter duration also brings computational benefits - since our model samples one random frame and audio segment during pre-training, we process only $30$-$40\%$ of the audio tokens compared to using full $10$-second segments. Audio segments shorter than $3$ seconds lead to degraded performance, suggesting this could be a limit for effective learning in our framework.

\noindent\textbf{Number of Frames. } Table~\ref{tab:ablation:frames} investigates the impact of temporal sampling density during both pre-training and evaluation. While increasing frames from $10$ to $16$ in either stage improves performance, the best results come from denser sampling in both, with pre-training and evaluating with $16$ frames yielding the best score. %

\noindent\textbf{Number of Registers. } In Table~\ref{tab:ablation:registers}, we assess the impact of varying register counts. While registers are known to boost classification performance in vision transformers \cite{darcet2024registers}, optimal counts can vary by task. We observe that increasing the number of registers consistently improves localization performance, from $20.9$ IoU with zero registers to $22.9$ with $16$ registers. 
We chose 8 register tokens as our standard setting since it achieved competitive performance across all tasks, supporting our goal of a balanced model.

\noindent\textbf{Global Token. } The presence of a global token is evaluated in Table~\ref{tab:ablation:cls}. Incorporating the global token enhances the performance for retrieval on AudioSet ($+3.6\%$ on average) and VGGSound ($+3.7\%$ on average), while also improving localization (mIoU from $20.7$ to $22.8$) and classification (AS20k mAP from $26.6$ to $30.8$). These consistent improvements across all tasks highlight the global token's role in capturing global context and helping disentangle the contrastive and reconstruction objectives.

\noindent\textbf{Masking Ratio. } Table~\ref{tab:ablation:masking} examines different masking ratios during pretraining. A lower masking ratio of $0.6$ achieves the highest retrieval performance but results in poorer localization (IoU $19.1$). A higher ratio of $0.9$ significantly degrades retrieval performance but improves localization (IoU $23.3$). A ratio of $0.75$ provides a good balance, with strong retrieval and localization (IoU $21.8$) performance. Applying a multi-ratio strategy \cite{lin2024avsiam} led to worse performance across all metrics. Therefore, we use $0.75$ masking in our final model, as it provides the best trade-off between retrieval and localization capabilities. %

\vspace{-2pt}
\section{Conclusion} \label{sec:conclusion}
In this work, we introduced CAV-MAE Sync, an extension of the popular CAV-MAE framework that addresses key challenges in audio-visual learning by treating audio as a temporally aligned sequence, disentangling contrastive and reconstruction objectives through separate global tokens, and enhancing spatial localization with learnable register tokens. 
Our experiments demonstrate across AudioSet, VGGSound, and ADE20K that these architectural improvements offer a more effective and efficient approach to audio-visual representation learning that harmoniously aligns temporal and spatial aspects of audio and visual modalities.

\vspace{-2mm}
\section*{\normalsize Acknowledgements}
\vspace{-2mm}
\small Edson Araujo is supported by German Federal Ministry of Education and Research (BMBF) project STCL - 01IS22067. This research is in part supported by the MIT-IBM Watson AI Lab.

{
    \small
    \bibliographystyle{ieeenat_fullname}
    \bibliography{main}
}

\clearpage
\setcounter{page}{1}
\maketitlesupplementary

\section{Implementation Details} \label{sec:implementation}

In this section, we provide details on our data preprocessing, model architecture, and training hyperparameters.

\subsection{Data Preprocessing}
For input, we sample 16 frames uniformly from each video, along with corresponding 4-second audio segments with temporal alignment determined as described in Section~\ref{sec:temporal_granularity}. For audio, each waveform is first converted to a sequence of $128$-dimensional log Mel filterbank (fbank) features computed with a $25$\,ms Hanning window every $10$\,ms., we extract $4$-second segments from the spectrograms of size $128\times416$, chosen to enable non-overlapping patch extraction. We use a patch size of $16\times16$, resulting in 208 audio tokens. The RGB images are resized and center-cropped to $224\times224$ pixels, following the same patch extraction process, resulting in 196 visual tokens. 

\subsection{Model Architecture}
For the model architecture, we initialize our modality-specific encoders from the same MAE checkpoints as CAV-MAE \cite{gong2023cavmae}, but conduct our own pretraining rather than using their pretrained weights. Our single-modality encoders each contain 11 transformer layers, followed by a 1-layer joint encoder for cross-modal fusion. This was chosen to maintain the compatibility with the original MAE architecture, from which we initialize our weights. For the linear probing downstream task, the final transformer classifier consists of 2 layers followed by a single linear layer applied to the $\text{CLS}$ token. 

\subsection{Training}
In all experiments we use a single backbone model pretrained on AudioSet2M \cite{gemmeke2017audio}. During pretraining, we use a masking ratio of $0.75$ for both modalities with unstructured masking following \cite{gong2023cavmae}. We conduct ablation studies on the impact of different masking ratios in Table \ref{tab:ablation:masking}. 

The contrastive and reconstruction loss weights are set to $\lambda_c=0.1$ and $\lambda_r=1.0$ respectively. For the contrastive loss weight $\lambda_c$, we use a higher value of $0.1$ compared to CAV-MAE's $0.01$, since aligning multiple fine-grained audio segments with their corresponding frames is a more challenging task than aligning a single global audio representation.

We use a batch size of $512$ and an initial learning rate of $2\times10^{-4}$ with cosine learning rate scheduling. We pretrain for $25$ epochs in total. Detailed hyperparameters for both pretraining and finetuning stages are provided in Table~\ref{tab:hyperpar}.

\begin{table}[t]
    \centering
    \resizebox{\linewidth}{!}{
    \begin{tabular}{@{}lcccc@{}}
    \toprule
                           & Pretraining & \hspace{5pt} & \multicolumn{2}{c}{Probing} \\ 
    \cmidrule(lr){2-2} \cmidrule(lr){4-5} 
    Dataset                & AS-2M   &    & AS-20K       & VGG  \\
    Optimizer              & \multicolumn{4}{c}{Adam, weight decay=$5\text{e}{-7}$, betas=($0.95$, $0.999$)} \\
    Learning Rate          & $2\text{e}{-4}$    &    & $5\text{e}{-2}$            & $1\text{e}{-3}$       \\
    LR Scheduler           & Cosine   &   & Cosine            & Cosine    \\
    Epochs                 & $25$      &   & $15$              & $10$        \\
    Linear Warmup Epochs   & $2.5$       &   & $1.5$               & $1$         \\
    Batch size             & $8\times64$   &    & $48$           & $48$        \\
    GPUs                   & $8\times\text{AMD MI200}$  &   & \multicolumn{2}{c}{$2\times\text{AMD MI200}$}               \\
    Training time          & $16\text{h}$     & & $2\text{h}$ & $10\text{h}$\\
    Audio Input Size       & $128\times416$  &  & $16\times128\times416$           & $16\times128\times416$\\
    Class Balance Sampling & No       &  & No                & Yes       \\
    Mixup                  & No      &   & Yes               & Yes       \\
    Random Time Shifting   & Yes     &   & Yes               & Yes       \\
    Loss Function          & -       &   & BCE                 & CE       \\
    Weight Averaging       & No      &   & Yes               & Yes       \\
    Input Norm Mean        & $-5.081$  &   & $-5.081$            & $-5.081$    \\
    Input Norm STD         & $4.485$   &   & $4.485$             & $4.485$     \\ \bottomrule
    \end{tabular}
    \vspace{-2mm}
    }
    \caption{Our pre-training and fine-tuning hyperparameters.}\label{tab:hyperpar}
\end{table}
\section{Modality-Specific Linear Probing} \label{sec:modality_specific_linear_probing}

Table~\ref{tab:linear-probing:modality} presents the results of our modality-specific linear probing experiments. We compare the performance of models trained with audio-only, video-only, and audio-visual inputs on two datasets: AudioSet-20K (AS20K) \cite{gemmeke2017audio} and VGGSound \cite{chen2020vggsound}. The results are reported using mean Average Precision (mAP) for AS20K and accuracy for VGGSound. The audio-visual model outperforms both the audio-only and video-only models, achieving the highest scores of $30.5$ mAP on AS20K and $52.7$ accuracy on VGGSound. This demonstrates the effectiveness of combining audio and visual modalities for classification tasks.

\begin{table}[ht]
    \footnotesize
    \centering
    \resizebox{0.7\linewidth}{!}{
    \begin{tabular}{@{}lccc@{}}
    \toprule
       \textbf{Modality}             & AS20K$\uparrow$ & VGGSound$\uparrow$  \\
    \midrule
    Audio Only & $8.7$  & $30.3$ \\
    Video Only & $22.3$ & $46.3$ \\
    Audio-Visual              & $\textbf{30.5}$      & $\textbf{52.7}$              \\
    \bottomrule
    \end{tabular}
    }
    \vspace{-2mm}
    \caption{Comparing audio-visual classification performance using linear probing. Numbers reported for AS20K are calculated using mAP and VGGSound with accuracy.}
    \label{tab:linear-probing:modality}
    \vspace{-5mm}
\end{table}

\section{Retrieval Aggregation Methods} \label{sec:retrieval_aggregation}
We evaluate different strategies for aggregating similarity scores in cross-modal retrieval, as shown in Table~\ref{tab:suppl:retrieval}. For any pair of videos, we compute a similarity matrix where each element represents the similarity between a visual token from the query video and an audio token from the target video, as detailed in Section~\ref{sec:method}. The ``diagonal mean'' strategy averages only the diagonal elements of this matrix, focusing on temporally aligned token pairs, while ``block mean'' averages all pairwise similarities between the two videos. Our experiments show that ``diagonal mean'' consistently outperforms other approaches, including ``block mean'' and maximum-based strategies (``diagonal max'' and ``block max''). This suggests that emphasizing temporal alignment through diagonal averaging better captures the audio-visual correspondences compared to considering all possible token pairs or focusing on single maximum similarity values. The advantage is particularly pronounced on AudioSet, where ``diagonal mean'' achieves $35.2\%$ R@1, surpassing ``block mean'' by $2.7\%$ and ``diagonal max'' by $6.7\%$ absolute.

\begin{table}[ht]
    
    \resizebox{\linewidth}{!}{
    \begin{tabular}{@{}cccccccc@{}}
    \toprule
    \multirow{2}{*}{Strategy} & \multicolumn{3}{c}{AudioSet Eval Subset} & \multicolumn{3}{c}{VGGSound Eval Subset} \\ 
    \cmidrule(lr){2-4} \cmidrule(lr){5-7}
    & R@1 & R@5 & R@10 & R@1 & R@5 & R@10 \\
    \midrule
    block max & $27.8$ & $51.9$ & $62.4$ & $23.0$ & $43.9$ & $54.3$ \\
    diag max & $28.5$ & $51.9$ & $61.3$ & $22.6$ & $43.5$ & $54.6$ \\
    block mean & $32.5$ & $54.8$ & $65.0$ & $25.9$ & $48.2$ & $59.2$ \\
    \rowcolor{almond}\textbf{diag mean} & $\textbf{35.2}$ & $\textbf{58.3}$ & $\textbf{67.6}$ & $\textbf{27.9}$ & $\textbf{51.7}$ & $\textbf{61.8}$ \\ 
    \bottomrule
    \end{tabular}}
    \vspace{-2mm}
    \caption{Comparison of retrieval aggregation strategies for cross-modal retrieval. The ``diagonal mean'' aggregation achieves the best performance, surpassing ``block mean'' by $2.7\%$ and ``diagonal max'' by $6.7\%$ absolute on AudioSet R@1. (V $\rightarrow$ A Retrieval.)} \label{tab:suppl:retrieval}
\end{table}
\vspace{-2mm}

\section{Register Tokens Analysis}

In this section, we analyze the information captured by different token types through linear probing on the AudioSet-20k dataset. Table~\ref{tab:register_analysis} shows the performance comparison between register tokens, patch tokens, and the global token. Our findings reveal that register tokens serve as an intermediate representation between highly localized patch tokens and the global token. 

With our proposed 8 registers setup, the global token achieves the highest performance ($30.8$ mAP), followed by register tokens ($17.8$ mAP) and patch tokens ($11.7$ mAP). This hierarchy indicates that register tokens effectively aggregate information from patches while maintaining more specialized representations than the global token. The performance gap between register and patch tokens ($17.8$ vs $11.7$ mAP) shows that registers capture more semantic information than individual patches.

Adding registers improves global token performance from $27.1$ to $30.8$ mAP, suggesting that registers serve as a "buffer" to aggregate information independently. Interestingly, we observe that register tokens reduce patch token performance from $12.3$ to $11.7$ mAP, indicating that registers are drawing contextual information away from patches. This supports our disentanglement hypothesis, while we don't observe the high-norm tokens reported in the paper that first introduced register tokens to vision transformers \cite{darcet2024registers}, this reduction in patch performance suggests registers are successfully serving as intermediaries between local and global representations. Our design uses registers to untangle the competing generative (patch tokens) and contrastive (global token) objectives. The empirical improvement in global token performance, coupled with the reduction in patch token performance, demonstrates that this additional buffer, not directly controlled by any loss, effectively helps the model develop more specialized representations.

\begin{table}[ht]
    \centering
    \resizebox{0.7\linewidth}{!}{
    \begin{tabular}{@{}cccc@{}}
    \toprule
    \multirow{2}{*}{\# Registers} & \multicolumn{3}{c}{AS20k (mAP) $\uparrow$} \\ 
    \cmidrule(lr){2-4}
    & Register & Patch & Global \\
    \midrule
    0 & N/A & 12.3 & 27.1 \\
    \rowcolor{almond}8 & 17.8 & 11.7 & 30.8 \\
    \bottomrule
    \end{tabular}
    }
    \vspace{-2mm}
    \caption{Linear probing of models with and without registers on AudioSet-20k, using various tokens as representation.}
    \label{tab:register_analysis}
\end{table}

\section{Sound Prompted Segmentation Examples} \label{sec:sound_prompted_segmentation}
\begin{figure*}[ht]
    \centering
    \includegraphics[width=\textwidth]{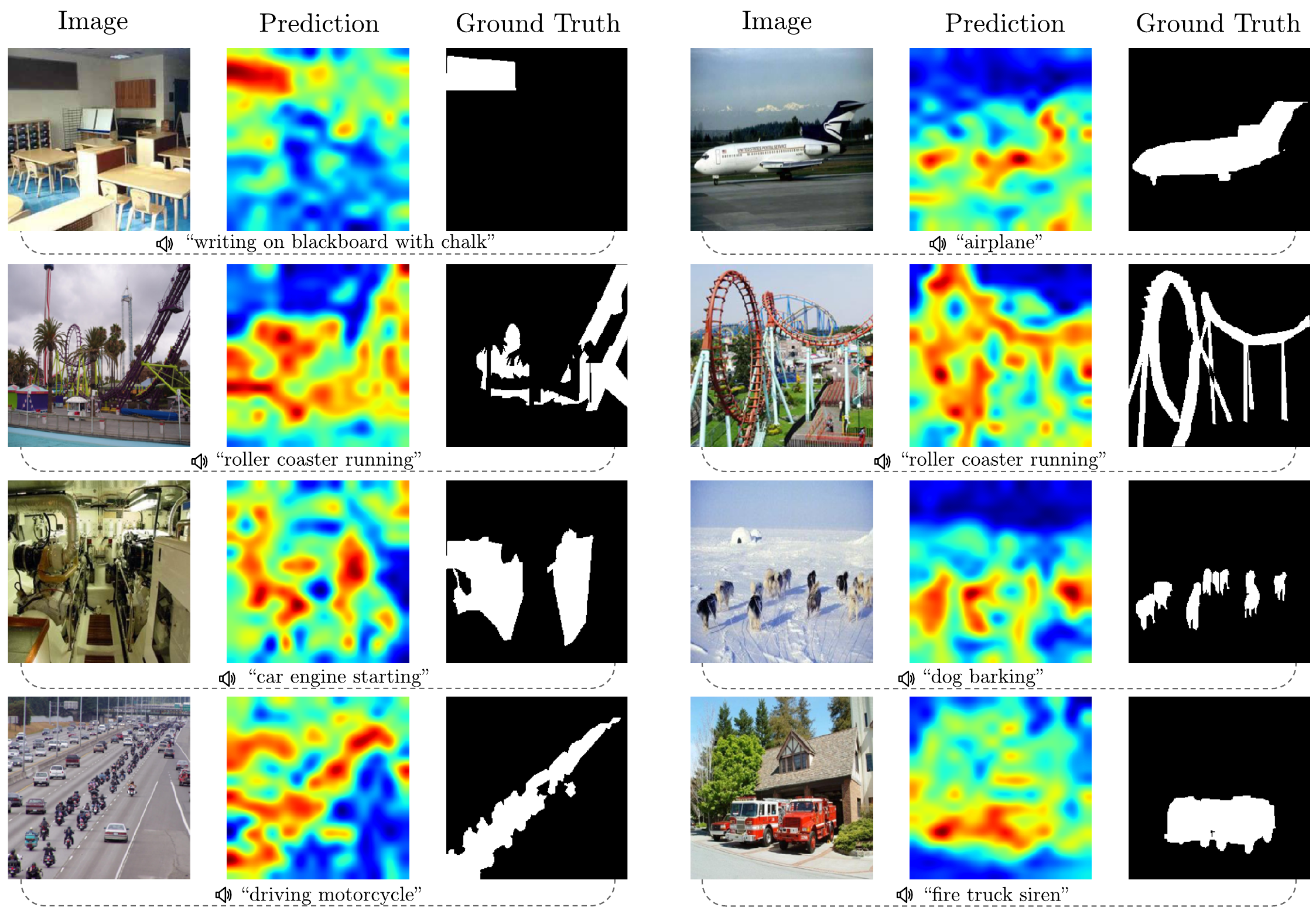}
    \caption{Sound-prompted segmentation results showing localization maps generated from audio prompts from VGGSound classes like "writing on blackboard", "roller coaster", and "airplane". The model highlights relevant image regions corresponding to the audio, demonstrating strong audio-visual alignment for clear objects while more complex, cluttered scenes remain challenging.}
    \label{fig:segmentation}
\end{figure*}

Figure~\ref{fig:segmentation} shows our model's sound-prompted segmentation results. As described in Section~\ref{sec:method}, we compute localization maps by calculating cosine similarities between the global audio token and visual patch tokens. Using VGGSound audios from class labels like ``writing on blackboard with chalk'', ``roller coaster running'', and ``airplane'' as prompts, our model generates localization maps highlighting relevant image regions. The results demonstrate strong audio-visual token alignment for scenes with clear objects like airplanes, also for more complex scenes like roller coasters with high visual clutter, which are naturally more challenging.

Notably, while specific classes like ``writing on blackboard with chalk'' and ``roller coaster running'' are not explicitly labeled in our AudioSet-2M pretraining dataset, examples of these sounds still exist under different labels. Despite this labeling discrepancy and the domain gap compared to datasets like VGGSound, our model demonstrates strong localization capabilities. For instance, in the "writing on blackboard" example, the model precisely highlights the blackboard region, while in the roller coaster examples, it effectively focuses on the coaster structure within visually cluttered scenes. These results are particularly encouraging given that our model was trained in a self-supervised manner on AudioSet-2M without explicit localization objectives. This robustness to unlabeled classes suggests that our global contrastive learning approach inherently learns some degree of spatial correspondences between audio and visual signals.

\section{Intra-Instance Temporal Segmentation} \label{sec:temporal_segmentation}
\begin{figure*}[ht]
    \centering
    \includegraphics[width=\textwidth]{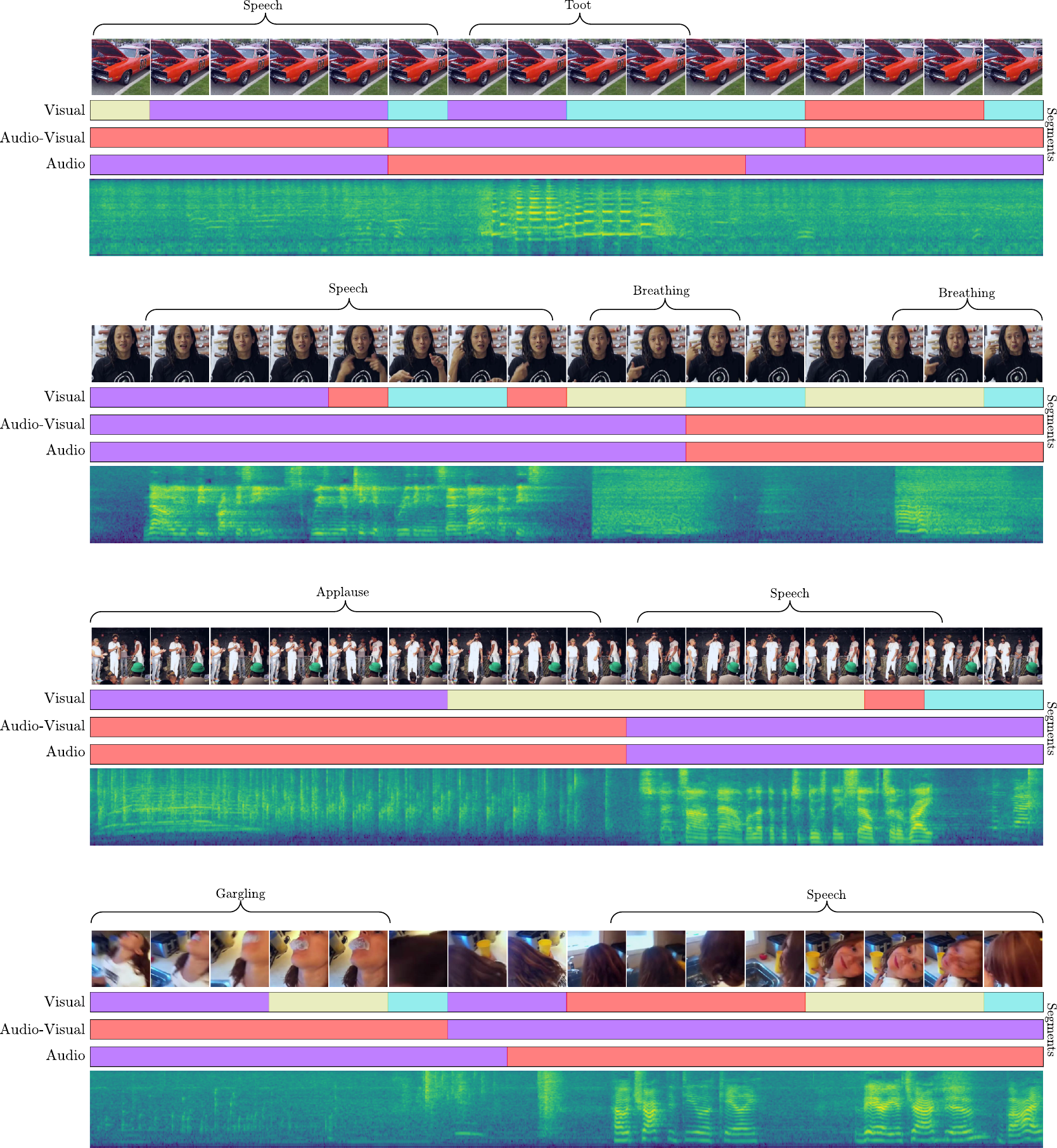}
    \caption{Temporal segmentation results showing audio-visual event boundaries across different scenarios. Each row shows frame sequences with corresponding segmentation bars for visual, audio-visual, and audio-only features, along with spectrograms. Labels indicate primary events (Speech, Text, Breathing, Applause, Gurgling) manually detected in different temporal segments}
    \label{fig:temporal_segmentation}
\end{figure*}

To investigate how finer-grained audio representations impact the understanding of video clips, we conduct a qualitative analysis of temporal segmentation within samples from the AudioSet dataset. For this experiment, we manually annotate the occurrence of the classes throughout the video. In many cases, especially when multiple classes are present, different classes occur in separate segments of a video, not necessarily overlapping. We observe how well our model's features can discern between these classes of audio events by extracting features from each of the 16 frames and corresponding audio segments.

We apply a simple adaptive clustering algorithm to the extracted features to create temporal segments within each video. Using Agglomerative Clustering with a dynamic distance threshold, we iteratively adjust the threshold to achieve the desired number of segments, which if set to $5$. If this fails, we fall back to K-means clustering. Figure~\ref{fig:temporal_segmentation} shows examples where our model can segment different classes based on the audio, even when the visual information remains nearly constant. We compare segmentation results using audio-only, video-only, and combined features to demonstrate how audio features capture most of the semantic changes occurring within videos. This highlights why using a single global audio representation would be insufficient, as it would fail to capture these important temporal variations in the audio signal.

In the first example with the red car, while the visual scene remains largely static, the model detects distinct "speech" and "toot" segments, demonstrating audio's ability to capture semantic transitions invisible in the visual domain. The second sequence shows clear delineation between speech and breathing segments, with audio features driving the segmentation despite minimal visual changes. The third example captures the transition from applause to speech in a crowd setting, where both audio and visual cues contribute to the boundary detection. The final sequence shows gurgling transitioning to speech, with audio features again providing the primary signal for segmentation.

Notably, the audio-visual segmentation (middle bar in each set) often closely matches the audio-only segmentation (bottom bar), suggesting that audio features frequently dominate the temporal boundary detection. This makes intuitive sense for events like speech, breathing, and applause that have distinct acoustic signatures but may not correspond to major visual changes.

These examples highlight the importance of processing audio in smaller segments rather than using a single global representation. The audio features are often more relevant for segmenting these videos, demonstrating the value of the fine-grained audio processing approach of CAV-MAE Sync.

\end{document}